\documentstyle[12pt,aasms4]{article}
\begin{document}

\title{Direct spectroscopic evidence for ionized 
       PAHs in the interstellar medium}

\author{G.C. Sloan$^1$~$^2$}
\authoremail{sloan@ssa1.arc.nasa.gov}

\author{T.L. Hayward$^3$}
\authoremail{hayward@astrosun.tn.cornell.edu}

\author{L.J. Allamandola$^2$}
\authoremail{lallamand@mail.arc.nasa.gov}

\author{J.D. Bregman$^2$}
\authoremail{jbregman@mail.arc.nasa.gov}

\author{B. DeVito$^3$}

\author{D.M. Hudgins$^2$}
\authoremail{hudgins@ssa1.arc.nasa.gov}

\affil{$^1$School of Physics, University College, Australian Defence 
           Force Academy, Canberra, ACT 2600, Australia\newline 
       $^2$NASA Ames Research Center, MS 245-6, Moffett Field, 
           CA 94035-1000\newline
       $^3$Center for Radiophysics and Space Research, Cornell
           University, Ithaca, NY 14853}

\begin{abstract}
Long-slit 8$-$13~$\mu$m spectroscopy of the nebula around
NGC 1333 SVS 3 reveals spatial variations in the strength and
shape of emission features which are probably produced by 
polycyclic aromatic hydrocarbons (PAHs).  Close to SVS 3, the 
11.2~$\mu$m feature develops an excess at $\sim$10.8$-$11.0~$\mu$m, 
and a feature appears at $\sim$10~$\mu$m.  These features 
disappear with increasing distance from the central source, and 
they show striking similarities to recent laboratory data of PAH 
cations, providing the first identification of emission features 
arising specifically from ionized PAHs in the interstellar medium.
\end{abstract}


\section{Introduction} 

Gillett, Forrest, \& Merrill (1973), in a spectroscopic study
of planetary nebulae, discovered a series of unidentified 
infrared (UIR) bands, now known to include features at 3.3, 
6.2, 7.7, 8.6, and 12.7~$\mu$m.  Since then the UIR bands have
been observed in a rich variety of astronomical sources 
(see Allamandola 1996 for a recent review).  Polycyclic 
aromatic hydrocarbons (PAHs) were first introduced as a 
possible carrier of the UIR bands by Leger \& Puget (1984) and
Allamandola, Tielens, \& Barker (1985).  While other carriers 
have been proposed for the UIR bands, none have proven as 
successful as the PAH model in explaining the observed 
spectral details.  Nonetheless, many differences exist between 
earlier laboratory and astronomical spectra, including the 
relative strength of the features and the wavelengths at which 
they appear, and these differences have prevented the PAH 
model from gaining wider adherence.  

A major discrepancy exists between the relative intensities 
of the features in the 10$-$13~$\mu$m region (C--H out-of-plane 
bending modes) and the features in the 6$-$9~$\mu$m region 
(C--C modes and the C--H in-plane bend at 8.6~$\mu$m).  In 
laboratory spectra of neutral PAHs, the C--C modes are much 
weaker than the features at longer wavelengths, but in 
astronomical spectra, the C--C modes are stronger.  Allamandola 
et al. (1985) originally suggested that most or all PAHs in 
the interstellar medium would be ionized, due to their low 
ionization potential ($\sim$6 eV), and laboratory 
investigations of PAH cations show better agreement with the 
observed UIR spectra, in terms of both the relative feature 
strengths and positions (Hudgins, Sandford, \& Allamandola 1994; 
Hudgins \& Allamandola 1995a, 1995b, 1997, 1998; Szczepanski 
\& Vala 1993a, 1993b; Szczepanski, Chapo, \& Vala 1993; 
Szczepanski et al. 1995a, 1995b).

Here we make a detailed comparison between recently available 
laboratory data and spatial and spectral variations in the 
infrared spectrum from the SVS 3 region in the reflection
nebula NGC 1333.  SVS 3 is an early B star (Strom, Vrba, \& 
Strom 1976; Harvey, Wilking, \& Joy 1984), producing a much
milder UV spectrum than found in other PAH emission regions
(e.g. the Orion Bar, NGC 7027).  Joblin et al. 
(1996) obtained discrete mid-infrared spectra in 3 locations 
in the reflection nebula using a 5$\arcsec$ beam and found 
that the strength of the 8.6 and 11.2~$\mu$m PAH features 
varied inversely to each other, with the 8.6~$\mu$m feature 
emitting more strongly near SVS 3 and the 11.2~$\mu$m 
feature emitting more strongly to the south away from SVS 3.  
Since the 8.6~$\mu$m feature shows enhanced strength in the 
spectra of most PAH cations (along with the other bands between 
6 and 10~$\mu$m), Joblin et al. concluded that SVS 3 was 
ionizing a larger fraction of the PAHs closer in than further 
away.  This dependence of ionization fraction as a function of
distance from the ionizing source could easily result from
geometric dilution.

This investigation concentrates on emission features in the
10$-$13~$\mu$m spectral region, which arise from out-of-plane 
bends of C--H bonds on the periphery of PAH molecules 
(e.g. Bellamy 1958; Allamandola, Tielens, \& Barker 1989).  
For {\it neutral} PAHs the strongest feature at 11.2~$\mu$m 
originates in PAH rings which contain only one C--H bond 
(the solo mode).  The duo mode (two adjacent C--H bonds) 
produces a feature in the vicinity of 11.9~$\mu$m, but this 
feature is much weaker in astronomical sources.  Since in 
the laboratory spectra of PAHs the wavelength of these bands 
can shift a substantial fraction of a micron, depending on 
the size and structure of the molecule, the positions of the 
solo and duo modes have long been used by chemists as a 
diagnostic of PAH structure.  The trio mode produces a feature 
at $\sim$13~$\mu$m, but the wavelength range of this mode 
overlaps with the quartet and quintet modes at longer 
wavelengths, making unambiguous identification difficult.  

\section{Observations and analysis} 

In order to investigate the spectral variations in the PAH emission 
at higher spatial resolution, we obtained long-slit 8$-$13~$\mu$m 
spectra of NGC 1333 SVS 3 at the 5-m Hale Telescope at Palomar on 
the nights of 1996 September 29$-$30 (UT) using SpectroCam-10 
(Hayward et al. 1993).\footnote{Observations at Palomar Observatory
were made as part of a continuing collaborative agreement between 
the California Institute of Technology, Cornell University, and the 
Jet Propulsion Laboratory.}  The data have a spectral resolution 
of 0.19~$\mu$m and a diffraction-limited angular resolution of
$\sim$0\farcs5.  The slit was oriented N/S and covered a
2$\times$16$\arcsec$ region of the sky including SVS 3 and the
nebulosity to the south. We used standard chop-and-nod sequences 
with 40$''$ and 60$''$ amplitudes E/W to correct for background
emission from the telescope and sky. The data were flux-calibrated
using spectra of $\beta$~Peg taken immediately before or after the
NGC 1333 observations, together with archival SpectroCam-10 ratio
spectra of $\beta$~Peg vs. $\alpha$~Lyr and the absolute 
$\alpha$~Lyr model from Cohen et al. (1992).  The data from the 
two nights were combined into a single 2-D spectral image from 
which individual 1-D spectra were extracted for plotting.

Figure 1 illustrates the spectrum summed from 2$\arcsec$ south of 
SVS 3 to the end of the slit, showing the 8.6~$\mu$m feature on 
the shoulder of the stronger 7.7~$\mu$m feature, the 11.2~$\mu$m 
feature and the emission plateau extending to the weaker 
12.7~$\mu$m feature.  Figure 2 shows how the strengths of the 
PAH features vary with position along the slit.  While the 11.2 
and 12.7~$\mu$m features grow progressively stronger toward the 
PAH emission ridge 10$\arcsec$ south of SVS 3, the 8.6~$\mu$m 
feature is stronger closer in.  This behavior confirms the 
results of Joblin et al. (1996). 

Fig. 3 shows how the shape of the 11.2~$\mu$m feature (solo 
mode) changes as a function of distance from SVS 3.  Close to 
the source, the feature has an excess on the short-wavelength
wing, which appears to consist of multiple components.  As the
distance from SVS 3 increases, the shorter wavelength portion
of the wing (centered at $\sim$10.8~$\mu$m) disappears first,
followed by the longer wavelength portion (at $\sim$11.0~$\mu$m).

We have extracted the flux from these two components (Fig. 4), 
which we describe as the ``blue outliers'' to the 11.2~$\mu$m
feature, by averaging the profile of the 11.2~$\mu$m feature 
8$\arcsec$ south of SVS 3 (and beyond), normalizing this mean 
profile to each row, and subtracting it.  The 10.8~$\mu$m outlier
is summed from 10.6 to 10.9~$\mu$m and the 11.0~$\mu$m outlier is 
summed from 10.9 to 11.2~$\mu$m.  While the 10.8~$\mu$m outlier 
goes to zero 8$\arcsec$ from SVS 3, the 11.0~$\mu$m outlier still 
makes a contribution, which we crudely estimate to be 
0.91$\times$10$^{-16}$ W m$^{-2}$ arcsec$^{-2}$ at this position 
by fitting a gaussian to the blue edge of the main band at 
11.2~$\mu$m.  Both outliers increase in strength by a factor of 
$\sim$2 close to SVS 3 despite the fact that the stronger features 
at 8.6, 11.2, and 12.7~$\mu$m are at a mininum in this region
(Fig. 2).

Our observations of NGC 1333 suggest that the 12.7~$\mu$m 
feature also develops a blue wing close to the central source.  
Because of the poor signal-to-noise in this spectral region, deep 
atmospheric absorption features (due to water vapor at 12.38, 
12.44, 12.52, and 12.56~$\mu$m and CO$_2$ at 
12.63~$\mu$m), and the complications introduced by possible 
[Ne II] emission at 12.78~$\mu$m, we cannot be more 
conclusive or quantitative with the present data.

An emission feature in the vicinity of 9.8~$\mu$m also appears
within 1$\arcsec$ of SVS 3 (Fig. 5).  This feature occurs on the 
wing of a very strong telluric absorption feature from O$_3$, 
making its apparent wavelength dependent on the quality of the 
atmospheric correction and difficult to determine accurately.  
However, the feature in our spectrum cannot arise entirely from 
a poor telluric correction, or it would appear along the entire
length of the slit and not just near SVS 3.  The 10~$\mu$m feature
also appears in the on-source spectrum of SVS 3 by Joblin et al. 
(1996), but without comment and only at the 1-$\sigma$ level.  
The feature has also appeared (faintly,  and without comment) in 
spectra obtained from the Infrared Space Observatory (Beintema et 
al. 1996) and the Infrared Telescope in Space (Yamamura et al. 
1996).  Both of these telescopes are above all atmospheric ozone.  
This feature probably does not arise from silicate dust, because 
silicates would produce a much broader emission band.

\section{Discussion} 

A component in the vicinity of 11.0~$\mu$m has appeared before 
in the spectra of several PAH sources, most strongly in TY CrA
(at $\sim$11.05~$\mu$m; Roche, Aitken, \& Smith 1991), but also
in Elias 1 (at $\sim$11.06~$\mu$m; Hanner, Brooke, \& Tokunaga
1994), and more weakly in several other sources, including Elias
14 (at $\sim$10.8~$\mu$m; Hanner, Brooke, \& Tokunaga 1995)
and WL 16 (DeVito \& Hayward 1998).  The WL 16 data (also
obtained with SpectroCam-10 at Palomar) show spatial behavior
similar to our NGC 1333 data, with the blue wing on the 
11.2~$\mu$m feature becoming more pronounced closer to the
central source and fading further away (Fig. 3 in DeVito \& 
Hayward 1998).

To determine the nature of the blue outliers and the 10~$\mu$m
feature, we compare our astronomical data to the database of
 spectra from the Astrochemistry Laboratory Group at NASA Ames 
Research Center (Hudgins et al. 1994; Hudgins and Allamandola 
1995a, 1995b, 1997, 1998) and theoretical models by Langhoff 
(1996).

For neutral PAHs, Langhoff (1996) finds that the position of the 
11.2~$\mu$m feature depends on the geometry of the molecule.  
Moving from a small molecule like anthracene (three adjacent 
rings) to tetracene (four rings) to pentacene (five rings), the 
feature shifts from 11.3 to 10.9~$\mu$m.   This shift raises the
possibility that the blue outliers might result from a change
in the composition of the PAH mixture.  However, the laboratory 
data of Hudgins et al. show a much smaller shift in wavelength as 
a function of molecular size.  From anthracene to pentacene, the 
feature only shifts from 11.3 to 11.1~$\mu$m.  

The relative strengths of the 11.2 and 12.7~$\mu$m bands provide 
a crude means of probing the size of the PAHs.  In larger PAHs, 
the solo mode (11.2~$\mu$m) would dominate, since any ring along 
a straight edge of the molecule would have only one C--H bond, 
while the trio mode (12.7~$\mu$m) would appear more frequently in 
smaller PAHs, where a larger fraction of the rings occupy corners
and not straight edges.  As Fig. 2 shows, the spatial behaviors 
of both the 11.2 and 12.7~$\mu$m features agree with each other 
within the uncertainties, pointing to closely related sets of 
carries and not variations in the size of the PAHs.  Consequently, 
we do not consider it likely that the blue outliers to the 
11.2~$\mu$m feature result from changes in the size distribution 
or molecular geometry of the PAHs sampled by the slit.

Figures 6 and 7 compare the laboratory and theoretical spectra 
of neutral PAHs and PAH cations.  In the cations, most of the 
C--H out-of-plane bending modes have shifted $\sim$0.4~$\mu$m 
to shorter wavelengths (Fig. 6), providing a straightforward 
interpretation of the blue outliers seen close to SVS 3.  Figure 
7 shows that the 10~$\mu$m feature seen near SVS 3 may also 
arise from PAH cations, since PAH cations consistently produce 
features in this wavelength region while neutral PAHs do not.

Joblin et al. (1996) argued that the fraction of PAH cations 
decreases further from SVS 3 due to decreasing fluxes of 
ionizing photons.  In our data, the blue outliers to the 
11.2~$\mu$m feature and the 10~$\mu$m feature show just this
spatial dependence.  The combination of this spatial behavior
and the appearance of similar spectral features in laboratory
spectra of ionized PAHs leads us to identify these features 
with PAH cations.

This identification substantially strengthens the case for
PAHs as carriers of the UIR bands.  The identification of the 
3.29~$\mu$m band with the aromatic C--H stretch and the
bands in the 11$-$13~$\mu$m region with out-of-plane C--H 
bends requires that the carrier consist of aromatic hydrocarbons,
but the nature of these aromatic hydrocarbon molecules has
remained in doubt (e.g. Sellgren 1994, Tokunaga 1997, Uchida
et al. 1998).  Energy requirements discussed in the proposal of
PAHs as possible carriers of the UIR bands (Leger \& Puget 1984; 
Allamandola et al. 1985) lead to estimates that these molecules
must contain $\sim$40$-$80 carbon atoms.  Therefore, they must be 
polycyclic.  But these molecules might exist within a larger 
matrix of hydrocarbons and other molecules, commonly 
described as hydrogenated amorphous carbon (HAC; see the 
recent review by Duley 1993).  In order for the molecules to be 
ionized, they {\it must be free molecules}, i.e. they must be 
separate from any HAC-like matrix.  In the face of these 
combined arguments, individual gas-phase PAHs must be the 
dominant emitters of the narrow UIR bands.

\acknowledgments

The authors thank Walt Duley and Craig Smith for helpful
discussions.  We would also like to express our gratitude to 
an anonymous referee who returned comments to us in only two 
weeks.  During the preparation of this manuscript, GCS was
supported by NSF grant INT-9703665 and graciously hosted by
the School of Physics, Australian Defence Force Academy, and
the Division of Physics and Electronics Engineering, University 
of New England.

\newpage

\figcaption{ 
The average spectrum from the PAH emission region south 
of SVS 3, determined from 2$\arcsec$ away from SVS 3 to the 
south end of the spectrometer slit.  The error bars  
represent 1-$\sigma$ uncertainties.}

\figcaption{ 
Equivalent fluxes from the PAH features at 8.6~$\mu$m
({\it top}), 11.2~$\mu$m ({\it middle}), and 12.7~$\mu$m ({\it 
bottom}), determined by fitting and subtracting a linear baseline 
from beneath each feature. The 11.2~$\mu$m spatial profile ({\it 
dashed line}) has been normalized and plotted ({\it top and 
bottom}) for comparison.}

\figcaption{ 
Normalized spectral profiles of the 11.2~$\mu$m feature after 
removing a linear baseline, color-coded by position.  The 
lengths of the vertical bars above the profiles illustrate the 
average 1-$\sigma$ uncertainty in the data for each spectral 
strip.  These strips are 1\farcs5 wide.}

\figcaption{ 
The strength of the blue outliers of the 11.2~$\mu$m feature as
a function of position in the slit, determined by fitting and 
subtracting the mean 11.2~$\mu$m feature 8$\arcsec$ south of SVS
3 and beyond.  The 10.8~$\mu$m outlier (summed from 10.6 to 
10.9~$\mu$m) is plotted with {\it squares}) and the 11.0~$\mu$m 
outlier (10.9$-$11.2~$\mu$m) is plotted with {\it circles}.  We
estimate that the 11.0~$\mu$m outlier still contributes 
$0.9\times10^{-16}$ W m$^{-2}$ arcsec$^{-2}$ at 8$\arcsec$ from
SVS 3.}

\figcaption{ 
The spatial behavior of the spectrum at 10~$\mu$m in 1$\arcsec$ 
steps, after removing a linear baseline.  The black error bars 
show how the uncertainties vary with wavelength over this spectral 
range, primarily as a result of absorption from ozone in Earth's 
atmosphere at 9.6~$\mu$m. These data have been smoothed in 
both the spectral and spatial directions.}

\figcaption{ 
Laboratory and theoretical spectra for comparison with Fig. 3, 
produced by generating gaussian profiles from the wavelengths 
and strengths published by Hudgins et al. (1994), Hudgins and 
Allamandola (1995a, 1995b, 1997) and Langhoff (1996). The spectra 
are the sum of anthracene, tetracene, and chrysene (chosen 
because these molecules produce emission features in the vicinity 
of 11.2~$\mu$m).}

\figcaption{ 
Laboratory and theoretical spectra for comparison to Fig. 5, 
generated as in Fig. 6, except that all molecules with both 
neutral and cation spectra in each set have been summed.}

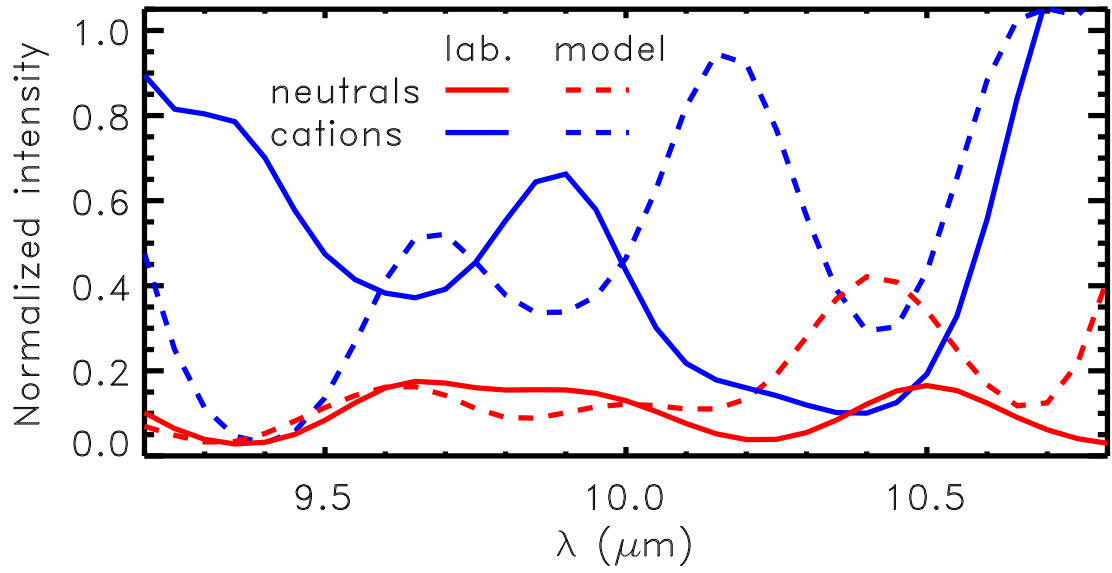
\end{document}